# *A la Recherche des Facteurs Déterminants de l'Intégration Internationale des Marchés Boursiers : une Analyse sur Données de Panel*

## AROURI Mohamed El Hedi

*EconomiX Université Paris X Nanterre*
*Bât G, 200, av. de la République Nanterre 92001 France*
*arourix@yahoo.fr*

### *Résumé*

L'objet de ce papier est d'identifier les facteurs déterminants de l'intégration financière internationale des marchés boursiers. En suivant un raisonnement intuitif, nous avons sélectionné un grand nombre de facteurs susceptibles d'agir sur l'intégration financière. Ensuite, nous avons développé un modèle international d'évaluation des actifs financiers à degré d'intégration variable au cours du temps. Ce modèle est alors estimé pour 30 pays (10 pays développés et 20 pays émergents) en faisant recours à l'économétrie des données de panel. Afin d'étudier si l'intégration financière des marchés émergents et celle des marchés développés réagissent différemment aux innovations économiques et financières, nous avons estimé le modèle aussi bien conjointement pour tous les marchés que séparément pour les marchés développés et les marchés émergents. Nos résultats montrent que l'ouverture au commerce mondial exerce un effet positif sur l'intégration financière de tous les marchés, que les facteurs globaux influent sur le niveau d'intégration des marchés développés alors que les facteurs liés à la stabilité économique et politique influent plutôt sur l'intégration financière des marchés des pays émergents.

### Determinants of International Stock Markets Integration: A Panel Data Analysis.

### *Abstract*

The aim of this paper is to identify the determinants of international stock markets integration. Intuitively we selected a great number of factors linked to financial integration. Then, we developed an international asset-pricing model with time-varying degree of integration. This model is estimated for 30 countries (10 developed countries and 20 emerging countries) using panel data econometrics. In order to investigate whether the financial integration in emerging markets and that in developed markets react differently to the economic and financial innovations, we estimated the model as well jointly for all markets as separately for developed markets and emerging markets. Our results show that trade openness exerts a positive effect on financial integration across all markets, the global factors drive integration in developed markets whereas the factors related to economic and political stability affect financial integration in emerging markets.

# 1- Introduction

La notion d'intégration financière a connu un intérêt grandissant ces dernières années. Son appréhension est utile aux entrepreneurs, aux investisseurs et aux opérateurs de marché. Pour les entrepreneurs, l'intégration financière présente notamment l'avantage de réduire le coût de capital grâce à un meilleur partage des risques, voir, entre autres, Bekaert et Harvey (2000) et Henry (2000). La baisse du coût de capital augmente la valeur actuelle nette des projets d'investissement et donc la profitabilité des entrepreneurs. Pour les investisseurs en portefeuilles, au fur et à mesure que le degré d'intégration financière internationale augmente, le poids relatif des facteurs globaux de risque devient plus important. Ainsi, l'intégration financière affecte aussi bien la structure de rentabilité que la structure de corrélations des actifs financiers. Il s'ensuit que les stratégies d'investissement en portefeuilles sont intimement liées au degré d'intégration des marchés financiers, voir entre autres, Longin et Solnik (2001), Karolyi et Stulz (2002) et Arouri (2004). De même dans le contexte actuel de globalisation et de concurrence accrue des places financières internationales, l'étude de l'intégration financière internationale intéresse les autorités économiques et monétaires. Pour être bref, il suffit de signaler que l'efficacité de toutes les actions de ces autorités dépend du niveau d'intégration financière.

Les travaux empiriques concernant l'intégration financière internationale sont nombreux, mais cette dernière demeure une variable difficile à définir, à mesurer et à en identifier les déterminants. Les études les plus récentes montrent que l'intégration financière est un processus dynamique évoluant au cours du temps en fonction des innovations économiques, financières et socio-politiques. Afin de pouvoir contrôler et tirer pleinement profit du processus d'intégration financière, il importe d'en identifier les déterminants. Malheureusement, il n'y a pas de théorie permettant d'identifier les facteurs affectant le degré d'intégration financière d'un marché. L'identification de ces facteurs incombe aux économistes économètres. Néanmoins, dans la littérature empirique des marchés financiers internationaux, on ne trouve pas de travaux ayant pour objectif l'identification des facteurs de l'intégration financière. Des travaux comme ceux de Bekaert et Harvey (1995,1997), Hardouvelis et al. (2002), Bhattacharya et Daouk (2002) et Adler et Qi (2003) ont arbitrairement retenu quelques variables financières et macroéconomiques dans la modélisation du degré d'intégration, mais n'ont pas formellement testé leur impact sur le degré d'intégration. Par ailleurs, dans ces travaux la notion d'intégration financière a toujours été étudiée d'un point de vue strictement individuel : le niveau d'intégration dépend des caractéristiques spécifiques à chaque pays. Or, l'intuition économique ne peut pas exclure l'existence des facteurs communs affectant conjointement l'intégration financière internationale de tous les marchés nationaux ou d'un groupe de marchés. Cet article vise l'identification à la fois des facteurs propres et des facteurs globaux à l'origine des variations spatiales et inter-temporelles des degrés d'intégration financière internationale.

L'identification des facteurs déterminants de l'intégration internationale des marchés boursiers nécessite la conception d'un modèle d'évaluation des actifs financiers permettant la distinction des sources systématiques globales et locales de risque. En effet, les marchés sont dits parfaitement intégrés si et seulement si les risques systématiques sont rémunérés de la même façon dans tous les pays. En d'autres termes, si les marchés sont complètement intégrés, les actifs ayant le même risque doivent avoir le même prix même s'ils sont traités sur des marchés différents. Sur des marchés parfaitement intégrés, les investisseurs courent des risques communs et des risques spécifiques à leurs pays ou secteurs d'activité, mais ils ne sont rémunérés que pour les sources communes de risque car les risques spécifiques sont totalement diversifiables. En revanche, si les marchés sont strictement segmentés, les investisseurs courent seulement des risques spécifiques. Dans ce cas, le même projet d'investissement dans deux pays différents peut avoir des rendements différents car les sources de risque et/ou leurs prix peuvent être différents. Toutefois, l'intégration parfaite et la segmentation stricte ne sont que des cas théoriques. En réalité, les marchés financiers se trouvent entre ces deux situations extrêmes, *i.e.* ils sont partiellement intégrés. Sur un marché partiellement intégré, les investisseurs courent et les risques communs et les risques spécifiques et le marché apprécie les deux sources de risque.

Dans la littérature des marchés financiers, les modèles d'évaluation des actifs financiers font l'une des deux hypothèses polaires suivantes : segmentation financière stricte ou intégration financière parfaite. Dans les modèles à segmentation stricte, les rentabilités anticipées des actifs financiers sont déterminées par les sources locales de risque. On peut par exemple citer les modèles domestiques de Sharpe (1964), Lintner (1965), Black (1972) et de Merton (1987). Dans le cadre international, les modèles d'évaluation supposent généralement que les marchés sont parfaitement intégrés. Dans ce cas, seules les sources internationales de risque sont pertinentes. Pour donner des exemples de modèles à intégration financière parfaite, on peut citer Harvey (1991), De Santis et Gérard (1997,1998), De Santis et al. (2003). Cependant, peu sont les travaux consacrés au cas plus réaliste des marchés partiellement segmentés.

La segmentation des marchés financiers a souvent été imputée à des facteurs d'ordres institutionnels: contrôle de capitaux, traitement fiscal, coûts de transaction, etc. Dans les modèles d'intégration financière partielle, ces barrières directes aux mouvements de capitaux ont souvent été traitées comme des coûts additionnels supportés par les investisseurs. Si ces coûts sont supérieurs aux gains attendus de la diversification internationale, les investisseurs préfèrent acquérir les actifs domestiques. On peut, entre autres, citer les travaux de Black (1974), Stulz (1981), Cooper et Kaplanis (1994,2000) et Errunza et Losq (1985,1992). Dans ces travaux, les modèles étudiés sont souvent mal-spécifiés car techniquement il est difficile voire impossible de tenir compte simultanément de plus de deux ou trois barrières. Or ces barrières directes aux mouvements internationaux de capitaux manifestent une très grande diversité inter-pays et inter-temporelle, voir Bekaert (1995). Toutefois, il importe de noter que ces barrières directes se sont considérablement réduites dans les deux dernières décennies par la suite aux vastes mouvements de déréglementation et de libéralisation qu'ont connus les marchés financiers, voir Bekaert et Harvey (2000) et Henry (2000). Néanmoins, les portefeuilles des investisseurs continuent à être déséquilibrés en faveur des actifs financiers domestiques, voir Lewis (1999). Si l'on croit les résultats des études les plus récentes, des barrières indirectes (risque de change, instabilités économiques et monétaires, risque pays, asymétries, etc.) privent les investisseurs globaux de profiter pleinement des opportunités offertes par les marchés étrangers.

Dans ce papier et dans l'objectif d'identifier les facteurs affectant le niveau d'intégration d'un marché boursier dans le marché international, nous développons un modèle international d'évaluation des actifs financiers à degré d'intégration variable au cours du temps en fonction des variables d'information liées à la conjoncture économique et financière nationale et internationale. Nous estimons ce modèle en faisant recours à l'économétrie des données de panel, *i.e.* séries individuelles-temporelles. L'utilisation des données de panel présente de nombreux avantages par rapport aux analyses traditionnelles en séries temporelles et en coupes instantanées. Une double dimension de l'information est disponible : une dimension individuelle (les marchés étudiés diffèrent les uns des autres) et une dimension temporelle (la situation de chaque marché varie d'une période à une autre). Ces données individuelles-temporelles constituent une source d'information extrêmement riche permettant d'étudier les phénomènes dans leur diversité comme dans leur dynamique. Estimer un modèle sur un panel accroît la taille de l'échantillon et apporte ainsi une richesse d'information importante comparativement à celle dont on dispose en économétrie des séries temporelles. En outre, les données de panel de pays permettent de rendre compte de l'influence des spécificités sociales, politiques et religieuses des pays (supposées non mesurables) sur leur de degré d'intégration financière internationale. En particulier, l'utilisation des données de panel permet d'enrichir considérablement la démarche économétrique et ce en testant la validité du modèle dans les dimensions individuelles et temporelles et en recherchant la conciliation des résultats obtenus dans ces deux dimensions.

Le reste du papier est organisé ainsi : la deuxième section présente un modèle international d'évaluation des actifs financiers à segmentation partielle avec déviations de la PPA ainsi que la méthodologie d'estimation retenue, la troisième section présente les données et quelques analyses préliminaires, la quatrième section expose les résultats empiriques et la cinquième section en tire les principales conclusions.

## 2- Méthodologie

Dans un premier temps, nous présenterons un modèle d'évaluation des actifs financiers à segmentation partielle et à degré d'intégration variable suivant les dates. Ensuite, nous exposerons la stratégie que nous allons emprunter afin d'estimer le modèle et étudier les facteurs déterminants de l'intégration internationale des marchés boursiers.

### 2.1- Le modèle

Dans un marché strictement segmenté, la prime de risque est déterminée par les facteurs locaux de risque. Dans le cadre du modèle d'équilibre des actifs financiers (MEDAF) de Sharpe (1964) et Lintner (1965), la prime de risque du titre (ou du portefeuille) *A* est reliée au risque du marché national *i*:

$$E(R_{it}^A - R_{ft} / \Omega_{t-1}) = \delta_d Cov(R_{it}^A, R_{it} / \Omega_{t-1}) \tag{1}$$

où $R_{it}^A$ la rentabilité du titre *A* (dans le pays *i*), $R_{it}$ est la rentabilité du portefeuille du marché du pays *i*, $R_{ft}$ est la rentabilité de l'actif sans risque et $\delta_d$ est le prix local de risque. Toutes les anticipations sont faites conditionnellement au vecteur informationnel $\Omega_{t-1}$ disponible aux investisseurs à l'instant (t-1). La prime de risque est alors égale au prix unitaire du risque domestique multiplié par l'exposition de l'actif considéré au risque du portefeuille du marché local.
Au niveau national, la relation (1) devient :

$$E(R_{it} - R_{ft} / \Omega_{t-1}) = \delta_d Var(R_{it} / \Omega_{t-1}) \tag{2}$$

Dans l'autre cas polaire d'intégration parfaite des marchés financiers internationaux, les rentabilités des actifs financiers sont générées par la version internationale conditionnelle du MEDAF :

$$E(R_{it} - R_{ft} / \Omega_{t-1}) = \delta_m Cov(R_{it}, R_{mt} / \Omega_{t-1}) \tag{3}$$

où $R_{mt}$ est la rentabilité du portefeuille du marché mondial et $\delta_m$ est le prix mondial de risque. Ainsi, si le marché est parfaitement intégré, la prime de risque attendue sur un actif donné est égale au prix unitaire du risque mondial multiplié par l'exposition de cet actif au risque du portefeuille du marché international.
Si on suppose en plus que la parité des pouvoirs d'achat (PPA) n'est pas vérifiée, le modèle doit inclure des primes liées au risque des taux de change. Dans ce cas, un modèle de type Adler et Dumas (1983) à *L+1* pays peut décrire les rentabilités attendues :

$$E(R_{i,t}^c / \Omega_{t-1}) - R_{f,t}^c = \delta_m Cov(R_{i,t}^c, R_{m,t}^c / \Omega_{t-1}) + \sum_{k=1}^{L} \delta_k Cov(R_{i,t}^c, R_{k,t}^c / \Omega_{t-1}) \tag{4}$$

où $R_k$ est la rentabilité du taux de change de la monnaie du pays *k* contre la monnaie du pays de référence *c*, $\delta_k$ est le prix du risque de change de la monnaie *k*. Toutes les rentabilités sont exprimées dans la monnaie du pays de référence *c*.

Toutefois, l'intégration parfaite et la segmentation stricte (intégration zéro) sont deux cas purement théoriques. En réalité, les marchés financiers nationaux expérimentent des cas intermédiaires, *i.e.* sont partiellement intégrés. Econométriquement, les deux cas extrêmes peuvent être combinés pour donner naissance au modèle à changement de régime probabiliste décrit par la relation (5) :

$$E(R_{i,t}^c / \Omega_{t-1}) - R_{f,t}^c = \psi_{i,t-1}\left(\delta_m\ Cov(R_{i,t}^c, R_{m,t}^c / \Omega_{t-1}) + \sum_{k=1}^{L} \delta_k Cov(R_{i,t}^c, R_{k,t}^c / \Omega_{t-1})\right) + (1-\psi_{i,t-1})\delta_d\ Var(R_{it} / \Omega_{t-1}) \quad (5)$$

où $\psi_{i,t-1}$ s'interprète comme une mesure conditionnelle du degré d'intégration financière du marché i dans le marché international. Le niveau d'intégration est supposé varier dans le temps. $\psi_{i,t-1}$ est compris entre 0 et 1. Si $\psi_{i,t-1}=1$, alors on est dans le premier régime où seuls les facteurs globaux de risque sont rémunérés. L'hypothèse de segmentation est alors rejetée. Si $\psi_{i,t-1}=0$, alors on est dans le second régime où seul le risque spécifique au marché i est rémunéré. Cette situation correspond à celle d'un marché strictement segmenté. Pour $\psi_{i,t-1} \in ]0,1[$, on est entre les deux régimes. Le marché étudié est donc partiellement intégré. Les rentabilités des actifs financiers y sont alors déterminées par une combinaison des facteurs globaux et locaux de risque.[1] Le degré d'intégration est modélisé grâce à une fonction de transition logistique :

$$\psi_{i,t-1} = \frac{\exp(\kappa_i Z_{i,t-1})}{1+\exp(\kappa_i Z_{i,t-1})}$$

où $Z_{i,t-1}$ est le vecteur de variables d'information sur le niveau d'intégration du marché i dont dispose les investisseurs à l'instant (t-1). Le choix de la forme logistique répond à une intuition économique claire: l'intégration financière nécessite au début beaucoup d'efforts. Une fois lancée, l'intégration devient souple et rapide. Toutefois, les marchés peuvent instantanément redevenir segmentés.

Enfin, notons que le modèle décrit par la relation (5) doit être estimé simultanément pour tous les pays étudiés. Toutefois, cette tâche s'avère très difficile étant donné que nous étudierons simultanément 10 pays développés et 20 pays émergents. Afin de réduire le nombre de paramètres à estimer, nous utilisons des indices composites des taux de change au lieu des taux de change bilatéraux. Dans la littérature des marchés financiers internationaux, les indices composites des taux de change ont été utilisés dans de nombreuses investigations empiriques. On peut, entre autres, citer les travaux de Ferson et Harvey (1993), Choi et al. (1998) et Carrieri et al. (2005). Nous utilisons deux indices composites: l'indice taux de change-pays développés et l'indice taux de change-pays émergents. Ces indices sont obtenus en fréquence mensuelle de la *Federal Reserve Bank of St Louis' FRED DataBase*.[2] Il s'ensuit que les primes de change obtenues sont des primes agrégées. En outre, afin de tenir compte du fait que contrairement à ce qui est souvent observé dans les marchés des pays développés, les variations des taux d'inflation dans les pays émergents ne sont pas toujours négligeables par rapport aux variations des taux de change, nous utilisons des indices des taux de change réels. Ce passage permet l'estimation du modèle aussi bien pour les marchés développés que pour les marchés émergents, voir Arouri (2005). En effet, seule l'inflation dans le pays de référence (Etats-Unis) est supposée non-aléatoire.[3]

---

[1] A chaque point du temps, le modèle (5) ressemble à la formulation statique du modèle à segmentation partielle de Errunza et Losq (1985) et de Cooper et Kaplanis (2000). Ces derniers interprètent $\psi_{i,t-1}$ comme la partie de la prime de risque expliquée par les facteurs globaux de risque.

[2] L'indice de change-pays développés est l'indice « *Major Paterns* » qui est une moyenne pondérée des taux de change de 16 pays développés : Allemagne, Australie, Autriche, Belgique-Luxembourg, Canada, Espagne, Finlande, France, Irlande, Italie, Japon, Pays-Bas, Portugal, Royaume-Uni, Suède et Suisse. L'indice de change-pays émergents est l'indice « O*ther Important Trading Paterns* » qui est une moyenne pondérée des taux de change de 19 pays émergents : Arabie-Saoudite, Argentine, Brésil, Chili, Chine, Colombie, Corée, Hong Kong, Inde, Indonésie, Israël, Malaisie, Mexique, Philippines, Russie, Singapour, Taiwan, Thaïlande et Venezuela. Ces indices sont disponibles en termes réels et nominaux. Pour plus de détails sur la construction de ces indices, se reporter à la « *Federal Reserve Bulletin*», octobre 1998, à l'adresse suivante : http://www.federalreserve.gov/pubs/bulletin/1998/1098lead.pdf.

[3] Dans la quasi-totalité des tests empiriques des modèles internationaux d'évaluation des actifs financiers, l'inflation domestique est supposée non-stochastique ou nulle. Ainsi, les taux de change nominaux sont utilisés afin d'approximer le terme inflation exprimée en monnaie de référence dans le modèle de Adler et Dumas (1983), voir par exemple Solnik

La figure 1 en annexes compare les indices de change réels et nominaux pour les deux groupes de pays. Pour les pays développés, les deux indices présentent des dynamiques similaires. Ce qui justifie le choix des études antérieures de supposer que les variations de l'inflation sont négligeables et de travailler directement sur des taux de changes nominaux. Cependant, pour les pays émergents, les deux indices présentent des différences significatives. Cela s'explique par le fait que l'indice taux de change–pays émergents inclut des taux de change des pays à fort taux d'inflation ayant expérimenté de nombreuses dépréciations de leur monnaie nationale. Supposer que l'inflation locale est négligeable et approximer les déviations de la PPA par les taux de change nominaux conduirait à des erreurs considérables et affecterait les résultats des estimations. Cela dit, nous utiliserons des indices de change réels. Le modèle à estimer s'écrit alors comme suit :

$$E(R_{i,t}^c / \Omega_{t-1}) - R_{f,t}^c = \psi_{i,t-1} \left( \delta_m \, Cov(R_{i,t}^c, R_{m,t}^c / \Omega_{t-1}) + \delta^d \, Cov(R_{i,t}^c, R_t^d / \Omega_{t-1}) + \delta^e \, Cov(R_{i,t}^c, R_t^e / \Omega_{t-1}) \right)$$
(6)
$$+ (1 - \psi_{i,t-1}) \delta_d \, Var(R_{it}^c / \Omega_{t-1})$$

où $R_t^d$ et $R_t^e$ sont les rentabilités des indices de change composites réels respectivement des pays développés et des pays émergents.

## 2.2- Méthodologie empirique

D'abord nous présentons la spécification économétrique, ensuite nous exposons la stratégie d'estimation que nous adopterons dans notre recherche des facteurs déterminants de l'intégration internationale des marchés financiers nationaux.

### 2.2.1- Spécification économétrique

Sous l'hypothèse d'anticipations rationnelles, le modèle décrit par (6) peut s'écrire :

$$R_{i,t}^c - R_{f,t}^c = \psi_{i,t-1} \left( \delta_m h_{imt} + \delta^d h_{idt} + \delta^e h_{iet} \right) + (1 - \psi_{i,t-1}) \delta_d h_{ii} + \varepsilon_t \; ; \; \varepsilon_t / \Omega_{t-1} \sim N(0, H_t) \quad (7)$$

où $h_{im,t}$, $h_{idt}$, $h_{iet}$ et $h_{ii,t}$ sont les colonnes de la matrice des covariances-variances $H_t$ mesurant respectivement les expositions au risque du marché mondial, au risque de l'indice des taux de change-pays développés, au risque de l'indice des taux de change-pays émergents et au risque du marché local. La matrice $H_t$ de taille $(N \times N)$ des variances-covariances est modélisée par le processus GARCH(1,1) multivarié asymétrique suivant :

$$H_t = C'C + aa' * \varepsilon_{t-1}\varepsilon'_{t-1} + bb' * H_{t-1} + ss' * \xi_{t-1}\xi'_{t-1} + zz' * \eta_{t-1}\eta'_{t-1} \quad (8)$$

où $C$ est une matrice triangulaire inférieure de taille $(N \times N)$, $\boldsymbol{a}$, $\boldsymbol{b}$, $\boldsymbol{s}$ et $\boldsymbol{z}$ sont des vecteurs de taille $(N \times 1)$ de paramètres constants
avec :
$\xi_{it} = \varepsilon_{it} I_{\xi_{it}}$ où $I_{\xi_{it}} = 1$ si $\varepsilon_{it} < 0$ et 0 sinon,
$\eta_{it} = \varepsilon_{it} I_{\eta_{it}}$ où $I_{\eta_{it}} = 1$ si $|\varepsilon_{it}| > \sqrt{h_{iit}}$ et 0 sinon.

---

(1974), De Santis et Gérard (1998), Hardouvelis et al. (2002), De Santis et al. (2003) et Phylaktis et Ravazzolo (2004). Toutefois, pour de nombreux pays émergents la variation de l'inflation locale est importante, ce qui complique davantage les tentatives de validation empirique du modèle de Adler et Dumas (1983). Le recours aux taux de change réels permet de surmonter cette difficulté, voir Arouri (2005) et Carrieri et al. (2005). En outre, le recours aux taux de change réels permet de dépasser certaines difficultés liées aux régimes des changes fixes fréquents dans les pays émergents.

La relation (8) implique que les variances dans $H_t$ dépendent asymétriquement du carré des innovations passées et d'un terme autorégressif, alors que les covariances dépendent asymétriquement du produit croisé des résidus passés et d'un terme autorégressif. En particulier, cette spécification garantit que la matrice des variances-covariances est définie et positive.

**2.2.2- Stratégie d'estimation**

L'estimation simultanée du modèle décrit par les relations (7) et (8) pour tous les marchés couverts par cette étude s'avère techniquement infaisable.[4] Dans notre recherche des facteurs déterminants de l'intégration financière internationale, nous adopterons, à l'instar de Bhattacharya et Daouk (2002), la démarche à deux étapes suivante :

*1$^{ère}$ étape*
D'abord, le système suivant est estimé pour chaque marché $i$ :

$$\begin{aligned}
R_{i,t}^c - R_{f,t}^c &= \alpha_i + \varepsilon_{it} \\
R_t^d &= \alpha_d + \varepsilon_{dt} \\
R_t^e &= \alpha_e + \varepsilon_{et} \\
R_{m,t}^c - R_{f,t}^c &= \alpha_m + \varepsilon_{mt}
\end{aligned} \qquad (9)$$

avec :
$$\varepsilon_t = (\varepsilon_{it}, \varepsilon_{dt}, \varepsilon_{et}, \varepsilon_{mt})' / \Omega_{t-1} \sim N(0, H_t)$$

et
$$H_t = C'C + A'\varepsilon_{t-1}\varepsilon_{t-1}'A + B'H_{t-1}B + S'\xi_{t-1}\xi_{t-1}'S + Z'\eta_{t-1}\eta_{t-1}'Z$$

L'estimation de ce système par la méthode du quasi-maximum de vraisemblance permet de récupérer pour chaque marché national $i$ la variance conditionnelle et les covariances conditionnelles avec le marché mondial et avec les taux de change.

*2$^{ème}$ étape*
Ensuite, une fois on dispose pour tous les marchés étudiés des séries des variances et des covariances, la relation (7) est alors estimée simultanément en données de panel pour tous les marchés en considérant successivement dans la fonction de transition $\psi_{i,t-1}$ l'un des facteurs candidats à l'explication des variations de l'intégration financière internationale. Le modèle à estimer est alors le suivant :

$$R_{i,t}^c - R_{f,t}^c = \psi_{i,t-1}\left(\delta_m h_{mi,t} + \delta^d h_{di,t} + \delta^e h_{ei,t}\right) + (1 - \psi_{i,t-1})\delta_d h_{ii,t} + \varepsilon_{i,t} \qquad (10)$$
$$i = 1,...,N, \quad t = 1,...,T_i$$

$$\psi_{i,t-1} = \frac{\exp(\kappa * facteur_{i,t-1})}{1 + \exp(\kappa * facteur_{i,t-1})}$$

Le modèle décrit par (10) constitue une extension du modèle de Bhattacharya et Daouk (2002) au cas où la PPA ne serait pas vérifiée et le risque des taux de change est internationalement rémunéré. Pour estimer ce modèle, la méthode des moindres carrés non-linéaires sera employée.

Cette deuxième étape permet d'identifier les facteurs qui contribuent significativement à l'explication de l'intégration financière internationale. Afin de pouvoir comparer les facteurs déterminants de l'intégration financière, la deuxième étape sera conduite en premier lieu conjointement pour tous les marchés et en second lieu séparément pour les marchés des pays développés puis pour les marchés des pays émergents.

---
[4] Pour les 31 marchés étudiés (30 marchés nationaux plus le marché mondial) on a 558 paramètres à estimer pour la matrice des covariances-variances.

Toutefois, le modèle de Bhattacharya et Daouk (2002) souffre d'une insuffisance majeure : le prix du risque domestique est supposé le même pour tous les pays. En d'autres termes, l'aversion au risque du portefeuille du marché domestique est la même pour tous les marchés nationaux indépendamment de leur degré d'intégration dans le marché mondial et de leurs spécificités individuelles. Cependant, des facteurs, souvent non-observables et/ou non-mesurables, peuvent affecter systématiquement l'aversion des investisseurs d'un marché donné au risque de leur portefeuille domestique. Il peut par exemple s'agir des préférences pour les produits nationaux, des réticences à certains instruments financiers pour de raisons religieuses et socioculturelles, de l'éloignement géographique, etc. Ces facteurs agissent sur l'aversion au risque domestique, sur le degré d'intégration du marché i et par conséquent sur la relation d'évaluation internationale donnée par le modèle (10). L'un des avantages essentiels des données de panel sur les autres types de données (séries temporelles et coupes instantanées) est de permettre la prise en compte de certaines caractéristiques individuelles non-observables. Afin de prendre en considération cette hétérogénéité des comportements des investisseurs représentatifs, nous autoriserons au prix du risque domestique de différer d'un marché national à l'autre, alors que le prix du risque mondial et les prix des risques des taux de change sont les mêmes pour tous les marchés étudiés. Le modèle ainsi défini est celui décrit par la relation (11) :

$$R_{i,t}^c - R_{f,t}^c = \psi_{i,t-1}\left(\delta_m h_{mi,t} + \delta^d h_{di,t} + \delta^e h_{ei,t}\right) + \left(1 - \psi_{i,t-1}\right)\delta_{di} h_{ii,t} + \varepsilon_{i,t} \qquad (11)$$

$$i = 1,...,N, \quad t = 1,...,T_i$$

L'introduction dans l'écriture du modèle (11) des variables indicatrices permet de rendre plus explicite le fait que les aversions individuelles aux risques locaux sont des coefficients à estimer :

$$R_{i,t}^c - R_{f,t}^c = \psi_{i,t-1}\left(\delta_{m,t-1} h_{mi,t} + \delta_{t-1}^d h_{di,t} + \delta_{t-1}^e h_{ei,t}\right) + \left(1 - \psi_{i,t-1}\right)\sum_{l=1}^{N}\delta_l \xi_{l,i,t} h_{ii,t} + \varepsilon_{i,t} \qquad (12)$$

$$i = 1,...,N, \quad t = 1,...,T_i$$

où :

$$\xi_{l,i,t} = \begin{cases} 1 & \forall t, l = i \\ 0 & \forall t, l \neq i \end{cases}$$

Dans le reste du papier, nous qualifions le modèle (10) de modèle sans effets individuels et le modèle (12) de modèle avec effets individuels.

## 3- Données et analyse préliminaire

En premier lieu, nous présentons les séries de rentabilités boursières des pays étudiés ainsi les séries des taux de change. En second lieu, nous présentons les facteurs déterminants éligibles de l'intégration financière internationale. En suivant un raisonnement intuitif s'inspirant de la théorie économique et des résultats des travaux antérieurs, nous discutons les liens qui peuvent exister entre ces facteurs et l'intégration financière internationale. Les statistiques descriptives de ces données sont reportées dans le Tableau 1.

### 3.1- Rentabilités boursières et taux de change

Cette étude couvre les marchés boursiers de 30 pays : 10 pays développés (Allemagne, Australie, Canada, Espagne, Etats-Unis, France, Italie, Japon, Royaume-Uni et la Suisse) et 20 pays émergents (Argentine, Afrique du Sud, Brésil, Chili, Corée, Egypte, Grèce, Hong Kong, Indonésie, Jordanie, Malaisie, Maroc, Mexique, Pologne, Singapour, Thaïlande, Turquie, Tunisie et Venezuela). Les observations utilisées sont les rentabilités mensuelles de fin de période de janvier 1973 à mai 2003.

Les données proviennent de *Morgan Stanley Capital International* (MSCI), de *International Finance Corporation* (IFC) et de *Datastream*. Les rentabilités boursières sont calculées avec réinvestissement des dividendes et en excès du taux des eurodollars à 30 jours issu de *Datastream*. Elles sont toutes exprimées en dollar américain.

Les marchés des pays émergents présentent des rentabilités en moyenne plus élevées que celles des pays développés. Toutefois, elles sont en moyenne nettement plus volatiles. Les coefficients d'asymétrie et d'aplatissement sont significatifs pour la quasi-totalité des marchés et l'hypothèse de normalité est rejetée. L'estimation de la version asymétrique du modèle par la méthode de quasi-maximum de vraisemblance rend compte de ces faits. Le test de Ljung-Box d'ordre 12 montre l'absence d'autocorrélation sérielle pour la plupart des séries boursières. Cependant, la première autocorrélation est significative pour la moitié des pays émergents étudiés.

Les statistiques descriptives des indices composites réels des taux de change issus de la *Federal Reserve Bank of St Louis' FRED DataBase* sont résumées dans le Panel B du Tableau 1. Les rentabilités moyennes sont égales à –0.19% pour l'indice pays-développés et 0.53% pour l'indice pays-émergents. L'hypothèse de normalité est rejetée pour les deux indices.

**3.2- Facteurs de l'intégration des marchés boursiers**

Dans cette sous-section, il est question de discuter et présenter les variables susceptibles d'être des facteurs déterminants propres et communs de l'intégration financière internationale. Comme nous l'avons déjà mentionné, la théorie économique ne permet pas l'identification de ces variables. Le choix des variables résulte des études antérieures et d'un raisonnement intuitif. Contrairement aux travaux antérieurs qui se sont arbitrairement limités à quelques facteurs susceptibles d'influencer le degré d'intégration d'un marché boursier donné (voir, entre autres, Bekaert et Harvey (1995,1997), Bhattacharya et Daouk (2002), Carrieri et al. (2003) et Adler et Qi (2003)), nous ne retenons pas de facteurs *a priori* et tous les facteurs sont éligibles. Ainsi, nous considérerons un grand nombre de facteurs susceptibles d'affecter le degré d'intégration internationale des marchés boursiers et nous testerons leur significativité. Les variables sélectionnées sont liées aux conjonctures économiques et socio-politiques nationales et internationales. En effet, de nombreux travaux aussi bien théoriques qu'empiriques montrent que les interactions entre le marché financiers et le reste de l'économie sont permanentes et que les marchés financiers réagissent rapidement et même anticipent les modifications de l'économie, voir notamment Prat (1982) et Fontaine (1987).

Pendant très longtemps, les barrières directes ont été considérées comme les principaux facteurs se trouvant dernière la segmentation financière. Par barrières directes on entend l'ensemble des barrières et contraintes imposées par les différents gouvernements tels que les taxes, les contrôles des capitaux, les coûts de transaction, etc. Cependant, nous pensons que le rôle de ces facteurs directs s'est considérablement estompé dans les dernières années et ce pour de nombreuses raisons. D'une part, les restrictions directes aux mouvements internationaux de capitaux se sont grandement réduites dans les deux dernières décennies. En fait, aussi bien les marchés développés que les marchés en développement ont connu des vastes mouvements de libéralisation. L'objectif était d'aller vers une plus grande ouverture des marchés financiers nationaux afin de tirer pleinement profit de l'intégration financière internationale (meilleure diversification des risques, réduction du coût de capital, croissance économique, etc.). D'autre part, de nombreuses études montrent que les investisseurs globaux peuvent toujours contourner les restrictions directes aux mouvements internationaux de capitaux en faisant notamment recours aux innovations financières et aux produits alternatifs, voir Bekaert (1995) et Glassman et Riddick (1996). Par ailleurs, pour affecter le degré d'intégration financière, ces barrières directes doivent être asymétriques c'est-à-dire elles doivent affecter de manière asymétrique les investisseurs nationaux et les investisseurs étrangers. Enfin, il faut signaler aussi que les barrières directes présentent une très grande hétérogénéité inter-pays et intertemporelle, ce qui rend difficile leur intégration dans un modèle d'évaluation des actifs financiers.

Toutefois, même après l'élimination progressive des barrières directes, certaines barrières indirectes peuvent persister et décourager les investisseurs étrangers. Bien qu'elles puissent être liées aux barrières directes, les barrières indirectes ne sont pas imposées par les gouvernements. Il peut par exemple s'agir des asymétries d'information, des instabilités macroéconomiques et des risques des changes. Si les nouvelles technologies d'information et de communication ainsi que la standardisation internationale des produits financiers et des normes comptables ont réduit considérablement tant les sources d'asymétrie d'information que les coûts de transaction et de traitement de l'information, les instabilités économiques et monétaires semblent s'intensifier avec la mondialisation. Nishiotis (2004) compare les effets des barrières directes et indirectes sur l'évaluation des fonds de pension. Il trouve que l'effet des barrières indirectes est grandement plus significatif que celui des barrières directes. Outre, ces barrières directes et indirectes, il ne faut pas ignorer le rôle qu'exercent les facteurs globaux sur l'architecture financière internationale. La restructuration financière internationale peut à son tour influencer l'intégration des marchés financiers nationaux.

Cela dit, dans ce travail, trois groupes de facteurs sont considérés : les facteurs locaux, les facteurs globaux et les mesures synthétiques de risque. Les statistiques descriptives de ces facteurs sont résumées dans le Panel C du Tableau 1 aussi bien conjointement pour tous les pays étudiés que séparément pour les pays développés et les pays émergents.

### *3.2.1- Les facteurs locaux*

Nous retenons ici des variables nationales susceptibles d'influer sur le niveau d'intégration internationale des marchés boursiers locaux. Ces variables sont censées refléter les cadres opérationnels et réglementaires, les niveaux de développement des marchés nationaux et les instabilités macroéconomiques et monétaires les caractérisant. Bien qu'elles soient propres aux marchés nationaux étudiés, ces variables sont le plus souvent liées à d'autres variables internationales. Toutefois, les coefficients de corrélation sont généralement faibles. Par exemple, Harvey et Ferson (1993) étudient les corrélations entre certaines variables internationales et des variables nationales des pays disponibles dans la base MSCI et trouvent des corrélations de moins de 40%.

#### *3.2.1.1- Degré d'ouverture commerciale*

Le ratio commerce extérieur (importations+exportations) sur PIB d'un pays donné traduit son intégration dans l'économie mondiale et son développement macroéconomique, voir notamment Henry (2000) et Carrieri et al. (2003). Une économie avec une bonne conduite économique et une bonne ouverture au commerce extérieur acquiert la confiance des décideurs et des investisseurs internationaux et attire plus de fonds étrangers. Ainsi, les économistes financiers pensent qu'une meilleure intégration économique conduirait à une meilleure intégration financière. Bekaert et Harvey (1997,2000) montrent que l'ouverture au commerce extérieur augmente l'exposition du marché local aux facteurs globaux de risque. Ainsi, plus le marché est ouvert au commerce extérieur, plus son degré d'intégration attendu est élevé. Nous retenons ainsi à l'instar de Bekaert et Harvey (1995,1997,2000), Rajan et Zingales (2000), Carrieri et al. (2003) et Bhattacharya et Daouk (2002) cette variable dans notre modélisation du degré d'intégration financière internationale. Notre hypothèse de base est la suivante : bien qu'elle suive un rythme différent, l'intégration économique favorise l'intégration financière.

Les données sont obtenues de la base de données *International Financial Statistics* (IFS) du Fonds Monétaire International. Le PIB est semestriel pour quelques pays et annuel pour d'autres. Pour dériver le PIB mensuel, nous divisons le PIB par 6 dans le premier cas et par 12 dans le second.

Le degré d'ouverture commerciale moyen est de 63.21%. Les pays développés sont en moyenne légèrement plus ouverts internationalement que les pays émergents (65.35% contre 61.08%). Le degré d'ouverture est plus volatile dans le cas des marchés développés que dans celui des marchés émergents.

#### *3.2.1.2. Développement du marché boursier*

Le développement du marché boursier est approximé par la variable capitalisation boursière sur PIB. Le développement du marché boursier est susceptible d'affecter les caractéristiques des rendements des titres à travers deux canaux. Le premier, le plus direct, est lié aux fonctions même du marché boursier qui doit faciliter les échanges, la diversification des risques, la fourniture de liquidité, le contrôle des firmes, etc. Plus la taille d'un marché augmente, plus sa capacité à mobiliser les fonds et diversifier les risques augmente. En outre, plus le marché est développé et liquide, plus les asymétries, en particulier l'asymétrie d'information, se réduisent, voir Levine et Zervos (1996). La réduction des asymétries, la liquidité et les possibilités de diversification des risques attirent les investisseurs étrangers, et donc agissent positivement sur le niveau d'intégration financière. Le second canal, indirect, passe par l'influence du marché financier sur les fondamentaux de l'économie qui affectent à leur tour la structure de rentabilités des titres financiers. Cette variable a été utilisée notamment par Bekaert et Harvey (1995,1997) et Carrieri et al. (2003) dans leur modélisation de degré d'intégration variable dans le temps, mais son effet sur le niveau d'intégration n'a pas été examiné. Le signe attendu sur cette variable est positif.

Cette variable est construite à partir des données issues de MSCI, IFC et DataStream. La moyenne du ratio capitalisation boursière sur PIB est de 26.35%. Les marchés des pays avancés sont en moyenne trois fois plus développés que ceux des pays émergents (35.10% contre 11.10%).

### *3.2.1.3- Flux de capitaux étrangers*

Les flux de capitaux étrangers reflètent en même temps la possibilité (barrières directes) et la volonté (barrières indirectes et facteurs globaux) des investisseurs globaux d'investir dans un marché donné. Ces flux de capitaux étrangers peuvent affecter la dynamique des marchés boursiers. Outre les phénomènes de contagion observés à court terme, l'ouverture à la prise de participation et au contrôle étranger a fait beaucoup pour promouvoir l'efficience du système financier et une gestion efficace des risques. On peut en outre penser que l'ouverture aux capitaux internationaux améliore d'autant plus l'efficacité du système financier qu'elle permet l'arrivée d'investisseurs étrangers (fonds d'investissement notamment). L'arrivée d'investisseurs et d'arbitragistes étrangers augmente la sensibilité du marché local aux facteurs globaux de risque et permettent une réduction des écarts des prix à la suite des opérations d'arbitrages. Bien évidemment, les possibilités d'arbitrage entre places financières ouvertes aux investisseurs globaux sont soumises aux contraintes définies par les dispositions légales et réglementaires en matière de mouvements de capitaux. De ce point de vue, l'intégration financière internationale est un processus graduel dont la vitesse dépend du cadre réglementaire et de la situation particulière de chaque marché national.

Nous considérons deux variables pour apprécier l'influence des actifs étrangers sur le processus d'intégration financière internationale : le ratio actifs étrangers sur PIB et le ratio actifs étrangers nets sur PIB. Ces données sont obtenues des bases IFS et *U.S Treasury International Capital*. Le ratio actifs étrangers nets sur PIB a une moyenne de 1.44%. Il est plus grand pour les marchés développés que pour les marchés émergents. Le ratio actifs étrangers sur PIB est égal en moyenne égal à 4.08%. Il est deux fois plus grand pour les marchés développés que pour les marchés émergents (5.22% versus 2.88%).

### *3.2.1.4- La volatilité des taux de change*

Selon certains économistes, la volatilité des taux de change constitue un frein au développement du commerce international et aux mouvements de capitaux, dans la mesure où elle rend très aléatoire les calculs économiques des entreprises importatrices ou exportatrices et des investisseurs globaux. Certes, il existe des techniques de couverture du risque de change qui permettent aux firmes et aux investisseurs effectuant des opérations à l'étranger de s'assurer contre une éventuelle variation du change. Mais ces techniques assurancielles ont évidemment un coût pour les entreprises et les investisseurs qui y ont recours, et sont donc susceptibles, au même titre que les variations de change elles-mêmes, de décourager des transactions internationales. Notre hypothèse de base est qu'une forte volatilité des taux de change peut entraîner la segmentation du marché boursier.

Les taux de change sont issus de la *Federal Reserve Bank of St Louis' FRED DataBase* et de IFS Plusieurs mesures de volatilité ont été retenues (la variance glissante, la variance des douze derniers mois et la variance issue d'une modélisation GARCH). Les résultats les plus significatifs sont ceux obtenus avec les modèles à effets GARCH(1,1). Seuls ces résultats seront reportés. La volatilité des taux de change bilatéraux avec le dollar américain est en moyenne de 0.27 (0.37 pour les marchés développés contre 0.19% pour les marchés émergents).

*3.2.1.6- Taux d'intérêt*
Nous considérons quatre variables : le taux d'intérêt local à court terme, la variation du taux d'intérêt court, la prime de terme et enfin la convergence du taux d'intérêt local vers les standards internationaux.

Les variations des taux d'intérêt à court terme affectent les mouvements internationaux de capitaux et peuvent donc agir sur le degré d'intégration financière internationale. La convergence des taux d'intérêt nationaux est utilisée dans les modèles macro-économique, comme mesure d'intégration financière internationale. Plus cette écart est faible, plus le niveau d'intégration est élevée. Quant à la prime de terme, elle a été utilisée par, entre autres, Hardouvelis et al. (2002) et Adler et Qi (2003) comme indicateur d'intégration des marchés financiers nationaux. Selon Adler et Qi (2003), cet écart mesure le risque souverain de chaque pays. Ce risque affecte l'engouement des investisseurs globaux et par conséquent les mouvements internationaux de capitaux. Ces mouvements de capitaux influent sur le degré d'intégration des marchés boursiers. Si l'écart des taux d'intérêt à terme augmente, le degré d'intégration diminue. Le signe attendu sur cette variable est donc négatif.

Les taux d'intérêt sont issus de la base IFS. Le taux d'intérêt à court terme présente une moyenne égale à 10.12% avec un écart-type de 12.03. Les taux d'intérêt sont clairement plus élevés pour les marchés de pays émergents que pour ceux des payés développés. La variation des taux d'intérêt courts à une moyenne de –0.05%, signalant une tendance générale à la baisse des taux d'intérêt. L'écart des taux d'intérêt courts par rapport à la moyenne des pays du G7 est égale à 4.22%. Cet écart est 7 fois plus grand dans le cas des pays émergents (8.19% versus 1.10%). La prime de terme est mesurée par l'écart entre le taux d'intérêt à court terme et le taux d'intérêt à long terme (souvent le rendement des obligations de l'Etat). Elle est en moyenne égale à 1.52%. Comme attendu, elle est plus grande pour les marchés émergents (2.69% versus 0.92%).

*3.2.1.6-L'inflation*
Les économistes s'accordent généralement à considérer que la stabilité monétaire et macro-économique est une condition importante pour le développement financier d'un pays. Outre la volatilité des taux de change bilatéraux, nous retenons deux autres variables : le taux d'inflation locale et la convergence des taux d'inflation. Selon Adler et Dumas (1983), Cooper et Kaplanis (1994) et Lewis (1999), les investisseurs investissent plus proportionnellement dans les actifs domestiques afin de se couvrir contre le risque d'inflation locale. Plus les taux d'inflation sont élevés, moins les investisseurs investissent internationalement. Ainsi, un lien négatif est attendu entre l'inflation et l'intégration financière internationale.

De même, la convergence des taux d'inflation locaux vers les taux internationaux incite les investisseurs à tirer profit des opportunités d'investissement offertes dans les marchés étrangers. Ces stratégies actives de diversification internationale des portefeuilles permettent de réduire les écarts de rentabilités ajustées des risques et donc favorisent l'intégration internationale des marchés boursiers. Ainsi, notre hypothèse est que le degré d'intégration financière internationale augmente avec la convergence des taux d'inflation. Nous considérons comme *benchmark* le taux d'inflation pondéré des pays industrialisés.

Enfin, notons aussi que nous étudions la significativité de la variation mensuelle du taux d'intérêt à court terme. Cette variable reflète les révisons d'anticipations de l'inflation locale, voir Gérard et al. (2003).

Le taux d'inflation est calculé à partir des indices des prix à la consommation issus de la base IFS. Le taux d'inflation moyen est de 8.48% (5.66% pour le pays développés versus 11.36% pour les marchés émergents). L'inflation est beaucoup plus forte et plus volatile dans les marchés émergents que dans les marchés développés. L'écart entre le taux d'inflation de chaque pays avec le taux d'inflation des pays industrialisés présente une moyenne annuelle de 2.85%. Cet écart est nettement plus faible dans le cas des pays développés (-0.20% versus 6.02%).

*3.2.1.7. Croissance de la production industrielle*
La production industrielle rend compte de l'état de l'activité du secteur industriel. Nous construisons deux variables : la croissance de la production industrielle et la convergence des taux de croissance nationaux vers les standards mondiaux.

En ce qui concerne la première variable considérée, de nombreux travaux tant théoriques qu'empiriques montrent l'existence d'un lien positif entre la croissance économique et la performance des marchés financiers. Ainsi, dans leur recherche de nouvelles opportunités d'investissement, les investisseurs globaux visent les pays qui enregistrent des taux de croissance élevés. La présence des investisseurs globaux sur un marché donné permet de réduire les écarts des prix et donc d'améliorer son intégration dans le marché mondial. Le signe attendu du lien entre la croissance de la production industrielle nationale et l'intégration financière internationale est donc positif.

Pour la deuxième variable retenue (convergence des taux de croissance), l'hypothèse que nous testons est très simple : si les marchés nationaux deviennent plus intégrés économiquement, les corrélations des cash-flows augmentent. En effet, dans la mesure où la synchronisation de ces mouvements pour plusieurs pays joue en faveur d'une synchronisation des mouvements de leurs marchés financiers (Phylaktis et Ravazzolo (2002)), une convergence des cycles économiques peut constituer un facteur sous-jacent à l'apparition d'une tendance commune dans l'évolution des prix des actifs financiers.

Pour étudier l'effet de cette synchronisation des cycles économiques, nous utilisons les indices mensuels de la production industrielle corrigés des effets saisonniers. Nous utilisons comme *benchmark* l'indice de production industrielle du G7. Nous calculons l'écart entre le taux de croissance de l'indice de production industrielle de chaque pays et celui du *benchmark*. Nous nous attendons à ce qu'une baisse de l'écart des taux de croissance se traduise par une augmentation du niveau d'intégration financière. Le signe attendu sur cette variable est donc négatif.

Nos variables sont construites à partir des données obtenues de la base IFS. La croissance moyenne de la production industrielle est de 0.98% (1.24% pour les pays développés contre 0.63% pour les pays émergents). L'écart moyen des taux croissance de la production industrielle est 0.10% (0.12% pour les pays développés contre 0.06% pour les pays émergents).

### *3.2.1.8- Autres facteurs*
Outre les facteurs précédents dont le lien avec les marchés boursiers a été mis en lumière par plusieurs travaux théoriques et empiriques, nous retenons les 3 facteurs suivants :

### *3.2.1.8.1- Le déficit courant*
Le déficit des comptes courants se traduit généralement par une entrée de capitaux finançant ce déficit. Cependant, la persistance d'un déficit entame la confiance des investisseurs. Le déficit courant moyen en pourcentage du PIB est –2.14% (-3.18% pour les marchés développés versus –1.10% pour les marchés émergents). Les données sont issues de la base IFS.

### *3.2.1.8.2- Le chômage*
Le taux de chômage est un indicateur économique qui s'est parfois révélé plus significatif que la production industrielle. Nous testons l'hypothèse suivante : la croissance du taux de chômage d'un pays donné est de nature à accroître le risque des titres de ce pays et donc à réduire son attractivité. Le singe attendu de cette variable est donc négatif.

Les données sont issues de la base IFS. Le taux de chômage moyen est égal à 7.54% (7.49% pour les pays développés versus 7.73 pour les pays émergents). Toutefois, il est utile de signaler que les taux de chômage ne sont pas disponible pour 75% des pays émergents couvets par cette étude.

### *3.2.1.8.3- La volatilité des exportations*
De nombreux travaux mettent en évidence le risque de change et le risque de défaut qui peuvent se déclencher dans les pays ayant une forte volatilité des exportations. Un pays dont les exportations sont volatiles fait recours aux capitaux étrangers pour lisser la consommation à travers les périodes de la variation du revenu afin de maintenir un bon rapport de crédit.

Les données sont issues de la base IFS. La volatilité moyenne des logarithmiques de exportations est égale à 10.85%. Les exportations sont plus volatiles dans les cas des pays émergents (6.97% pour les marchés développés versus 13.67 pour les marchés émergents).

### *3.2.2- Les facteurs globaux*

Outre les facteurs spécifiques aux marchés étudiés présentés dans la sous-section précédente, nous étudierons l'effet de quelques facteurs globaux sur le niveau d'intégration internationale des marchés boursiers nationaux. On entend par facteur global d'intégration tout facteur susceptible d'influencer conjointement l'intégration financière de plusieurs marchés nationaux. De nombreux travaux empiriques ont montré que les facteurs globaux affectent l'évaluation internationale des actifs financiers notamment pour les marchés développés, voir par exemple Harvey et Ferson (1993). Néanmoins, les travaux antérieurs n'étudient pas explicitement le lien entre ces facteurs globaux et l'intégration financière.

### *3.2.2.1- Taux intérêt internationaux*

Les travaux empiriques et théoriques montrent que dans le cas des marchés ouverts, les taux d'intérêt internationaux affectent l'évaluation des actifs financiers et les mouvements internationaux de capitaux, voir entre autres, Chinn et Forbes (2003) et Kose et al. (2003). Ces taux d'intérêt affectent l'allocation internationale des capitaux et motivent les opérations d'arbitrage international. Une baisse des taux d'intérêt internationaux augmente les mouvements de capitaux vers les marchés des pays émergents, voir par exemple Bekaert, Harvey et Lumsdaine (2002). On s'attend alors à une relation négative entre l'évolution des taux d'intérêt mondiaux et l'intégration financière : plus les taux d'intérêt internationaux sont faibles, plus les investisseurs globaux accèdent aux marchés des pays émergents et plus leur niveau d'intégration dans le marché international augmente.

Nous considérons deux variables : le taux d'intérêt international et les variations du taux d'intérêt international. Nous prenons comme *proxy* du taux d'intérêt mondial, le taux d'intérêt pondéré des pays du G7. Les données proviennent de la base IFS. Notre *proxy* du taux d'intérêt mondial a une moyenne annuelle de 6.505% avec un écart-type de 2.91. La variation du taux d'intérêt mondial est de –0.01% en moyenne.

### *3.2.2.2- La croissance mondiale*

De nombreux travaux récents établissent des liens entre la croissance économique mondiale et la performance des marchés financiers. D'une part, une croissance économique mondiale soutenue augmente le moral des investisseurs et leur confiance dans les marchés boursiers et donc permet une meilleure mobilisation internationale de l'épargne. La recherche d'opportunités profitables d'investissements intensifie les arbitrages internationaux et conduirait une meilleure intégration financière internationale. De ce point de vue, un lien positif est attendu entre la croissance mondiale et l'intégration financière internationale. D'autre part, une faible croissance de la production industrielle des pays développés augmente les mouvements internationaux de capitaux en destination des pays émergents. De ce point de vue, la relation entre la croissance dans les pays industrialisés et l'intégration des marchés boursiers des marchés émergents est négative. Nous approximons la croissance économique mondiale par la croissance de l'indice mensuel de production industrielle des pays industrialisés. Les données sont issues de la base IFS. Notre *proxy* de la croissance mondiale présente une moyenne de 0.137% avec un écart-type de 0.642.

### *3.2.3- Mesure synthétique de risque : le risque pays*

L'analyse du risque-pays est incontournable dans un contexte de mondialisation particulièrement suite à l'intensification accrue des phénomènes de rupture dans la sphère financière émergente et l'indissociabilité entre risque et investissement. Selon Clei (1998), le risque-pays peut être appréhendé comme étant *"l'ensemble des paramètres –macroéconomiques, financiers, politiques et sociaux – qui peuvent contribuer à la formation d'un risque autre que strictement commercial lors d'une opération avec un pays émergent".* De ce fait, le risque-pays peut englober deux composantes interdépendantes: d'une part, une composante « risques politiques et sociaux », résultant soit d'actes ou de mesures prises par les autorités publiques locales ou du pays d'origine, soit d'événements internes ou externes ; d'autre part, une composante « risques économiques et financiers », qui recouvre aussi bien une dépréciation monétaire qu'une absence de devises se traduisant, par exemple, par un défaut de paiement.[5] Ainsi, la notation risque-pays passe par l'évaluation de multiples critères d'origine économique, financière, politique et sociale. Plus généralement, le risque-pays désigne souvent la probabilité de survenue d'événements modifiant le résultat des entreprises et la valorisation des actifs dans un pays. Erb et al.

---
[5] Voir aussi Marois (1990).

(1996) illustrent ce type d'analyse et confirment que le risque-pays est pris en compte sur les marchés des actions des économies émergentes.

Dans le présent travail, les notations risque-pays sont obtenues de *l'Institutional Investor's Survey of Bankers*. A partir d'un panel d'une centaine de banquiers, *Institutional Investors* fournit des scores entre 100 et 0 en tenant compte des différents éléments d'incertitude qui se matérialisent par une volatilité spécifique du retour sur investissement international par rapport à un investissement domestique (défaut, suspension, rééchelonnement de paiement ; dévaluation, inconvertibilité des monnaies locales ; répudiation, contrôle des capitaux, illiquidité, corruption, risque politique, etc). Généralement, cinq catégories sont distinguées : risque très élevé (0-49.5), risque élevé (50-59.5), risque moyen (60-69.5), risque faible (70-84.5) et risque très faible (85-100). Ces données, à quelques exceptions près, commencent à partir de septembre 1979. Dans les estimations, nous utilisons le logarithme de ces indices à l'instar de Erb et al. (1996). Enfin, notons qu'un risque pays élevé décourage les investisseurs étrangers. On s'attend alors à ce qu'un niveau élevé de risque pays (indice faible) s'accompagne d'un faible degré d'intégration financière. Le signe attendu sur cette variable est donc positif.

Les risque pays en mars 2005 des marchés couverts par cette étude sont représentés dans la Figure 2. Cette figure montre bien que ce risque est plus élevé dans les pays émergents. Ceci est confirmé par les statistiques descriptives. La moyenne du risque pays est de 65.88. Les pays émergents sont de loin plus risqués que les pays développés (49.82 pour les pays émergents versus 86.11 pour les pays développés). De plus, les indices risque pays présentent une plus grande disparité dans le cas des pays émergents (un écart-type de 16.12 pour les pays émergents contre 8.36 pour les pays développés).

**4- Résultats**

Les résultats des estimations sont reportés dans le Tableau 2. Pour chaque facteur de l'intégration, les modèles sont estimés aussi bien conjointement pour tous les pays que séparément pour les pays émergents et les pays développés. Comparativement aux méthodologies retenues par les études antérieures, cet exercice permet d'étudier si les marchés émergents et les marchés développés réagissent différemment aux variations des facteurs susceptibles d'affecter leur degré d'intégration financière internationale. Les résultats de l'estimation du modèle avec effets individuels (relation (12)) sont résumés dans le Panel A du Tableau 2 et ceux de l'estimation du modèle sans effets individuels (relation (10)) sont reportés dans le Panel B du même tableau.

Les estimations du prix du risque du portefeuille du marché international et des prix des risques des taux de change et des marchés domestiques ne sont pas reportés dans le Tableau 2.[6] Le prix du risque mondial est significativement positif dans la quasi-totalité des cas étudiés. Ce qui montre que le facteur risque du portefeuille du marché mondial contribue significativement à l'explication des primes de risque dans les différents marchés. Les prix des risques des taux de change sont significatifs dans la plupart des cas. Ce résultat confirme les résultats des études précédentes et montre que les risque de change est rémunéré aussi bien pour les marchés développés que pour les marchés émergents. Des valeurs significatives des prix des risques domestiques sont trouvées notamment dans le cas des pays en développement couverts par cette étude. Ce résultat suggère la présence d'une segmentation partielle dans certains marchés.

Nous nous intéressons au signe et à la significativité de chacun des facteurs déterminants potentiels de l'intégration financière internationale. Commençons par le degré d'ouverture commerciale. Le développement des relations commerciales, vecteur d'intégration économique internationale, représente une source significative d'intégration financière. Le signe positif obtenu pour le degré d'ouverture commerciale est conforme à l'intuition économique : l'intégration financière est positivement liée à l'intégration économique. Cette variable est significative à 1% aussi bien pour tous les marchés que pour les marchés émergents et développés pris séparément. Le même résultats est obtenu dans les deux spécifications du modèle (avec ou sans effets individuels). Ce résultat confirme ceux d'autres investigations empiriques. Par exemple, Carrieri et al. (2003) et Bhattacharya et Daouk (2002) reportent des liens significatifs entre l'intégration financière internationale et l'intégration économique approximée par le degré d'ouverture commerciale. Chinn et Forbes (2003) montrent que le commerce bilatéral contribue significativement à l'explication des co-mouvements des marchés des actions et des obligations. Enfin, Bracker et al. (1999) étudient les co-mouvements bilatéraux de 9 marchés boursiers nationaux en utilisant une méthode d'estimation en deux étapes. Ils trouvent que ces co-mouvements bilatéraux sont affectés par certaines variables macroéconomiques liées à la compétitivité internationale et à l'ouverture commerciale des pays étudiés.

---

[6] Pour le modèle avec effets individuels, on a un prix du risque mondial, 2 prix des risques de change et 30 prix des risques domestiques pour chaque facteur de l'intégration. Pour le modèle sans effets individuels, on a un prix du risque mondial, deux prix des risques de change et une moyenne des prix des risques domestiques pour chaque facteur de l'intégration. Ces prix des risques ne sont pas reportés dans le Tableau 2 pour des raisons de clarté de l'exposé des résultats.

La variable développement du marché boursier approximée par le rapport capitalisation boursière sur PIB a un signe positif. La taille d'un marché améliore sa liquidité et sa capacité à mobiliser les fonds et diversifier les risques. Ainsi, un marché bien développé a plus de chance d'avoir un bon niveau d'intégration financière internationale. Toutefois, le coefficient de cette variable n'est statiquement différent de zéro aux niveaux conventionnels de risque que pour les marchés émergents dans le cas du modèle avec effets individuels. Carreiri et al. (2003) utilisent une méthode différente et trouvent que cette variable est positivement corrélée à un indice pré-estimé d'intégration financière internationale.

Considérons maintenant les flux de capitaux étrangers. Les variables actifs étrangers et actifs étrangers nets ont les signes attendus suggérant que l'intégration financière est positivement liés aux parts des titres étrangers. Cependant, les coefficients obtenus sur ces variables ne sont pas statistiquement différents de zéro exception faite du cas des pays émergents pour le modèle sans effets individuels à 10%. Dans leur étude des effets de la libéralisation des marchés financiers, Bekaert, Harvey et Lumsdaine (2002) trouvent que cette dernière augmente les mouvements nets des capitaux en moyenne seulement dans les trois premières années qui suivent l'ouverture des marchés. L'effet de cette ouverture s'estompe à long terme. Nos résultats suggèrent que les mouvements de capitaux constituent un mauvais *proxy* de l'intégration internationale des marchés de capitaux. Dans les investigations empiriques antérieures, les résultats sont mixtes. Par exemple, Bordo et Helbling (2003) trouvent que les flux financiers n'affectent pas le degré de synchronisation des cycles économiques. Par contre Imbs (2003) et Kose et al. (2003) trouvent que ces flux internationaux des capitaux étrangers influent positivement sur les co-mouvements des cycles économiques. Enfin, Lane et Milesi-Ferretti (2003) montrent que le ratio actifs étrangers sur PIB est significativement corrélé au degré d'ouverture économique et financière.

La volatilité des taux de change reflète généralement des incertitudes se reportant à la situation économique du pays en question et à l'efficacité des actions entreprises ou envisagées. Si le coût de couverture du risque des taux de change est supérieur aux gains attendus de la diversification ou n'est pas complètement compensé par la prime de change, les investisseurs globaux n'accèderont pas au marché considéré, ce qui conduirait à sa segmentation. Selon Hardouvelis et al. (2002), une forte volatilité des taux de change restreint les investisseurs à investir plus généreusement dans leurs marchés domestiques, et agit ainsi comme une barrière aux mouvements internationaux des capitaux. Les résultats de nos estimations confirment cette hypothèse. En effet, pour les deux spécifications retenues du modèle (avec ou sans effets individuels), le coefficient de la volatilité des taux change est négatif et significatif à 10%. Quand on étudie séparément les pays développés et les pays émergents, on trouve que la volatilité des taux de change n'est pas significative pour les marchés des pays développés et qu'elle est significative à 1% pour les marchés émergents. Ainsi, la volatilité des taux de change paralyse l'évolution de l'intégration financière internationale des marchés émergents. Ces résultats confirment ceux de Bodart et Reding (1999) et Bracker et al. (1999) qui étudient le rôle des taux de change dans les co-mouvements des marchés financiers et trouvent que les corrélations entre ces marchés dépendent négativement de la variabilité des taux de change.

Pour le facteur inflation, les signes obtenus sont mixtes mais sont tous non significativement différents de zéro. En ce qui concerne la convergence des taux d'inflation, le signe obtenu est positif indiquant une corrélation positive entre la convergence des taux d'inflation et l'intégration financière internationale. Cependant, les coefficients obtenus ne sont pas statistiquement différents de zéro.

Considérons maintenant le facteur taux d'intérêt. Le signe obtenus pour les pays développés est négatif, mais non significatif. Pour les marchés émergents, on obtient un signe positif indiquant que la hausse des taux d'intérêt augmente leur niveau d'intégration internationale. Toutefois, les coefficients obtenus ne sont pas significatifs aux niveaux habituels de risque. La variation des taux d'intérêt court est significativement positive à 10% pour les marchés émergents et développés étudiés conjointement dans le cas du modèle avec effets individuels, mais elle n'est pas statistiquement différente de zéro dans les autres cas. Contrairement à ce que nous attendions, le coefficient de la prime de terme est positif pour tous les marchés. Cependant, la prime de terme n'est en aucun cas significative. L'écart des taux d'intérêt, souvent utilisé comme mesure d'intégration financière internationale, ne semble avoir aucun effet significatif sur le degré d'intégration des marchés boursiers nationaux.

En ce qui concerne la croissance de production industrielle, le coefficient de cette variable est négatif mais non significatif pour les marchés développés. Pour les pays émergents, le coefficient de cette variable est positif suggérant qu'une forte croissance économique dans un pays émergent attire les investisseurs étrangers et améliore le niveau d'intégration de son marché. Toutefois, cette variable n'est significative que dans le cas du modèle sans effets individuels et ce à 10%. Récemment, Dumas, Harvey et Ruiz (2003) montrent pour certains marchés émergents l'existence d'une relation positive significative entre les corrélations des taux de croissance des PIB et celles des marchés financiers. En ce qui concerne la convergence des taux de croissance de la production industrielle, aucun résultat significatif n'a été obtenu.

Un déficit des comptes courants se traduit généralement par une entrée de capitaux étrangers et donc devrait agir positivement sur l'évolution de l'intégration financière internationale. Toutefois, ce déficit s'il persiste entame la confiance des investisseurs et peut entraîner une sortie massive de capitaux et donc agir négativement sur l'intégration financière. Les résultats de nos estimations montrent un lien positif entre le déficit courant et l'intégration financière internationale. Cependant, ce lien n'est pas significatif aux niveaux conventionnels de significativité.

Dans le cas du modèle avec effets individuels, la variable chômage n'est pas significative aussi bien pour tous les marchés étudiés conjointement que pour les marchés développés et émergents étudiés séparément. Dans le cas du modèle sans effets individuels, le facteur chômage est négativement significatif à 10% pour tous les marchés et pour les marchés de pays développés. Ainsi, la hausse du chômage détend le climat socio-économique et affecte négativement l'intégration financière des marchés nationaux.

Le facteur volatilité des exportations présente un signe négatif indiquant un lien négatif entre les fortes variabilités des exportations et l'intégration financière internationale. Toutefois, ce lien n'est pas significatifs aux seuils conventionnels.

Considérons maintenant les facteurs globaux de risque. Le taux d'intérêt mondial a un signe négatif. Ce qui suggère que l'intégration financière internationale est une fonction décroissante des taux d'intérêt mondiaux. Une baisse des taux d'intérêt dans les pays industrialisés amène les investisseurs à chercher des opportunités d'investissements plus rentables et donc accroisse les mouvements internationaux de capitaux, ce qui conduirait finalement à réduire les écarts de rémunération entre les différents marchés. Cependant, ce facteur global de risque n'est pas significatif pour les marchés émergents bien qu'il ait le signe attendu. Quant à la variation du taux d'intérêt mondial, on trouve pratiquement des résultats similaires. Ces résultats confirment ceux de Chinn et Forbes (2003) et Kose et al. (2003) qui ont reporté que les taux d'intérêt internationaux jouent un rôle significatifs dans l'explication des co-mouvements des marchés financiers des pays développés.

La croissance de la production industrielle des pays développés a un signe positif suggérant ainsi que la croissance économique mondiale favorise l'intégration financière internationale. Toutefois, cette variable n'est significative que dans un seul cas sur six : le cas des pays développés pour le modèle sans effets individuels.

Pour le risque pays, le coefficient estimé est positif. Selon Oetzel et al. (2001) et Borio et Packer (2004), le risque pays est le résultat de l'interdépendance des sphères d'ordre économique, financier et politique. Un niveau élevé de l'indice correspond à un marché à faible risque pays, ce qui attire les investisseurs étrangers et devrait donc conduire à une meilleure intégration. Cette hypothèse semble bien être vérifiée. En effet, dans le cas du modèle avec effets individuels, le risque pays est significativement positif à 10% pour tous les marchés étudiés conjointement et pour les marchés des pays émergents. Ces résultats confirment ceux des travaux empiriques antérieurs bien que ces derniers n'aient pas explicitement étudié le lien entre le risque pays et l'intégration financière. Par exemple, Bekaert (1995) et Nishiotis (2004) montrent le risque pays influe sur les mouvements internationaux de capitaux. Erb et al. (1996) et Bilson et al. (2002) trouvent que le risque politique permet d'expliquer une partie de la structure de rentabilités des actifs financiers dans les marchés des pays émergents. Toutefois, bien qu'il ait le signe pertinent, le risque pays n'est pas significatif dans le cas du modèle sans effets individuels.

Pour conclure, les facteurs globaux influent plutôt sur le degré d'intégration financière internationale des marchés des pays développés, mais ils n'ont pas d'effets significatifs sur l'intégration des marchés des pays émergents. Cette dernière est plutôt affectée pour les facteurs locaux liés à la stabilité économique et politique notamment la volatilité des taux de change bilatéraux et le risque pays. En outre, l'ouverture commerciale internationale favorise l'intégration financière de tous les pays.

## 5- Conclusion

Dans ce papier, nous avons adopté une démarche intuitive et simple afin d'identifier les facteurs déterminants de l'intégration financière internationale. Dans un premier temps, nous avons identifié les variables qui pouvaient nous renseigner sur le degré d'intégration. Ensuite, nous avons développé et estimé un modèle international d'évaluation des actifs financiers à degré d'intégration variable au cours du temps en fonction des variables d'information liées à la conjoncture nationale et internationale. Nous avons estimé ce modèle en faisant recours à l'économétrie des données de panel. L'utilisation des données de panel présente notamment les trois avantages suivants : accroître la taille de l'échantillon, tenir compte des effets des spécificités individuelles sur le degré d'intégration financière et tester la validité du modèle d'évaluation dans les dimensions individuelles et temporelles. Nous avons considéré 30 marchés boursiers : 10 marchés développés et 20 marchés émergents. Afin d'étudier si les marchés émergents et les marchés développés réagissent différemment aux innovations économiques et financières, nous avons estimé le modèle aussi bien conjointement pour tous les marchés que séparément pour les marchés développés et les marchés émergents.

Nos résultats principaux peuvent se résumer dans les points suivants. Le risque du portefeuille du marché mondial et les risque des taux de changes sont significatifs pour la quasi-totalité des marchés étudiés, alors que le risque local n'est significatif que pour quelques marchés émergents. L'ouverture au commerce mondial exerce un effet positif sur l'intégration financière de tous les marchés. Les facteurs globaux influent sur le niveau d'intégration des marchés développés. Les facteurs liés à la stabilité macro-économique et politique influent sur l'intégration des marchés financiers des pays émergents. Dès lors, si les pays émergents veulent tirer pleinement profit du processus d'intégration financière internationale des leurs marchés, ils doivent accompagner la libéralisation de ces derniers par des changements structurels visant plus de stabilité économique, monétaire et politique. Ces mesures doivent améliorer l'attractivité de ces pays comme destinations d'investissements étrangers.

# Références

# Annexes

# Figure 1 : Indices de change

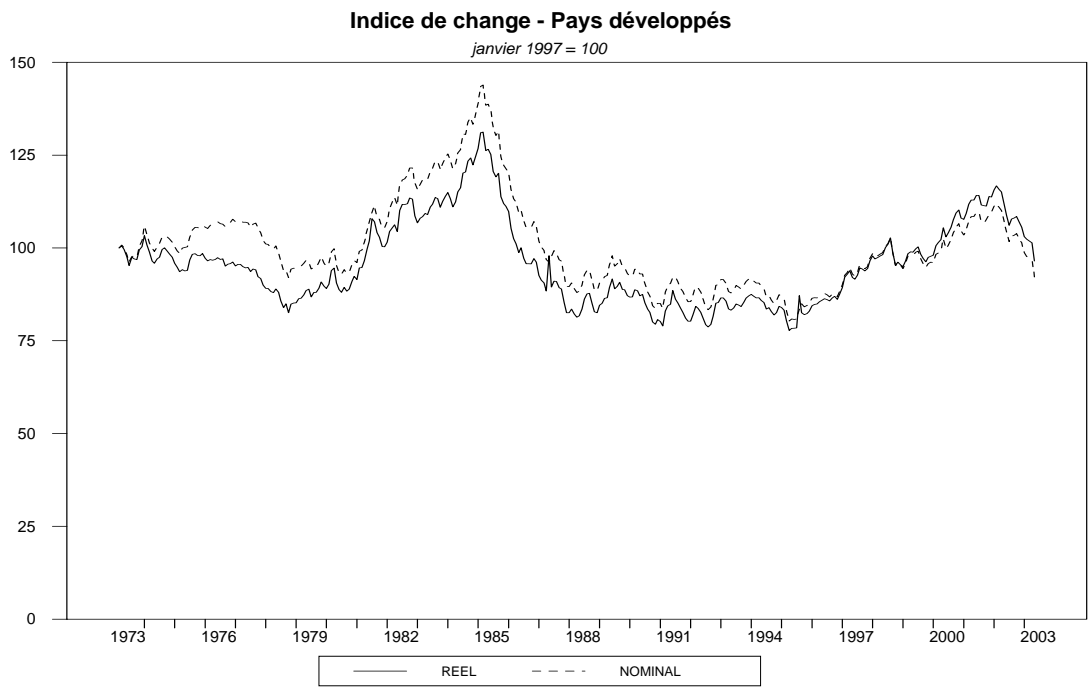

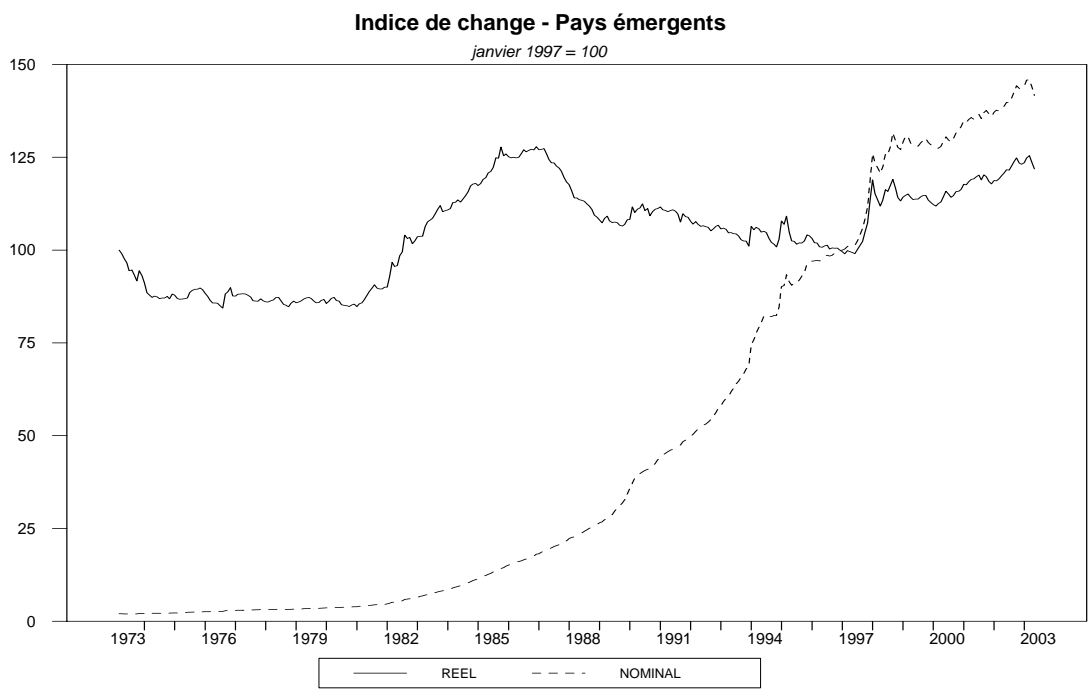

**Figure 2 : Risque Pays – mars 2005**.

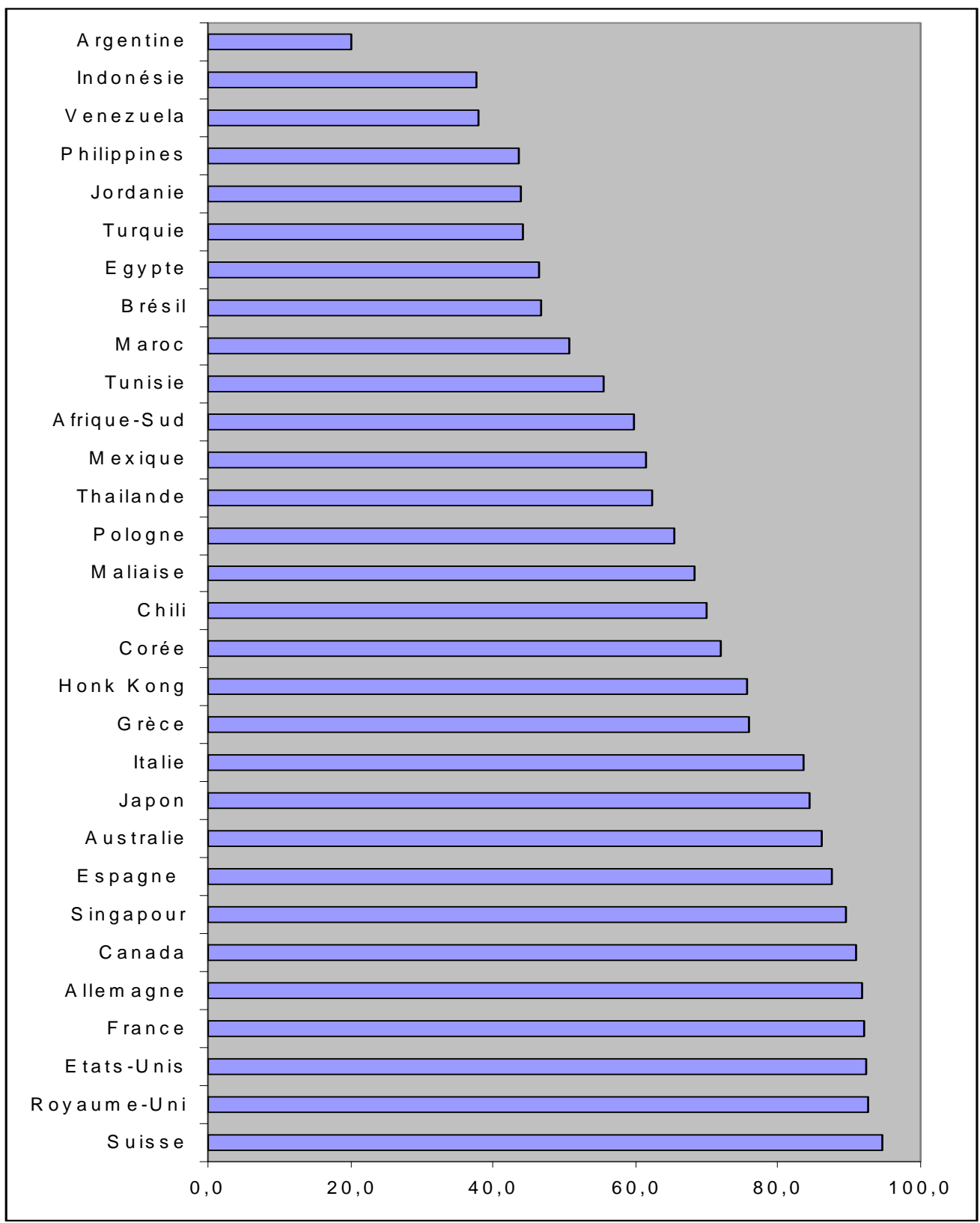

# Table 1
# Statistiques descriptives

**Panel A- Séries de rentabilités boursières**

**Pays développés**

|  | Début | Moyenne | Ecart-Type | Skewness | Kurtosis | J.B. | 1ère auto. | Q(12) |
|---|---|---|---|---|---|---|---|---|
| **Allemagne** | 1973 :01 | 5.678 | 76.618 | -0.222 | 1.415* | 33.462* | -0.028 | 15.629 |
| **Australie** | 1973 :01 | 5.509 | 86.227 | -0.802* | 5.136* | 440.394* | -0.029 | 5.976 |
| **Canada** | 1973 :01 | 3.449 | 68.607 | -0.429* | 1.835* | 62.428* | 0.028 | 12.664 |
| **Espagne** | 1973 :01 | 4.244 | 81.974 | 0.007 | 1.399* | 29.794* | 0.063 | 24.603** |
| **Etats-Unis** | 1973 :01 | 4.927 | 55.665 | -0.279** | 1.662* | 46.763* | 0.001 | 10.815 |
| **France** | 1973 :01 | 6.23 | 81.189 | -0.034 | 1.301* | 25.804* | 0.055 | 9.454 |
| **Italie** | 1973 :01 | 4.639 | 93.457 | 0.248 | 0.528** | 8.016* | 0.043 | 17.099 |
| **Japon** | 1973 :01 | 5.993 | 79.271 | 0.316** | 0.581* | 11.201* | 0.070 | 22.168** |
| **Royaume-Uni** | 1973 :01 | 6.619 | 82.794 | 1.414* | 11.597* | 2167.303* | 0.083 | 13.159 |
| **Suisse** | 1973 :01 | 6.613 | 66.993 | -0.048 | 1.334* | 27.214* | 0.066 | 9.211 |

**Pays émergents**

*Amérique latine*

|  | Début | Moyenne | Ecart-Type | Skewness | Kurtosis | J.B. | 1ère auto. | Q(12) |
|---|---|---|---|---|---|---|---|---|
| **Argentine** | 1988 :01 | 15.679 | 213.661 | 1.940* | 8.137* | 626.490* | 0.052 | 5.954 |
| **Brésil** | 1988 :01 | 27.794 | 207.322 | 0.404** | 3.481* | 98.471* | -0.120** | 15.748 |
| **Chili** | 1988 :01 | 14.446 | 90.176 | -0.006 | 1.130* | 9.856* | 0.187* | 16.700 |
| **Mexique** | 1988 :01 | 21.361 | 122.032 | -0.372** | 0.985* | 11.754* | 0.074 | 17.356 |
| **Venezuela** | 1993 :01 | 14.085 | 181.011 | 0.426 | 2.533* | 37.215* | -0.191* | 16.331 |

*Asie*

|  | Début | Moyenne | Ecart-Type | Skewness | Kurtosis | J.B. | 1ère auto. | Q(12) |
|---|---|---|---|---|---|---|---|---|
| **Corée** | 1988 :01 | 6.246 | 146.904 | 1.249* | 5.642* | 293.494 | 0.008 | 10.832 |
| **Hong Kong** | 1973 :01 | 10.694 | 133.341 | 0.991* | 10.978* | 1892.637* | 0.062 | 17.921 |
| **Indonésie** | 1988 :01 | 13.617 | 200.098 | 1.853* | 8.801* | 703.004* | 0.133* | 23.666** |
| **Malaisie** | 1988 :01 | 5.912 | 116.754 | 0.605* | 4.590* | 173.729* | 0.127** | 24.917** |
| **Philippines** | 1988 :01 | 1.344 | 121.355 | 0.556* | 2.260* | 48.913* | 0.242* | 20.176 |
| **Singapour** | 1973 :01 | 3.266 | 104.149 | 0.308** | 5.153* | 409.730* | 0.058 | 16.429 |
| **Thaïlande** | 1988 :01 | 6.372 | 148.347 | 0.206 | 1.479* | 18.170* | 0.040 | 31.413 |

*Europe Orientale*

|  | Début | Moyenne | Ecart-Type | Skewness | Kurtosis | J.B. | 1ère auto. | Q(12) |
|---|---|---|---|---|---|---|---|---|
| **Grèce** | 1988 :01 | 11.668 | 133.130 | 1.618* | 5.589* | 318.166* | 0.033 | 18.612 |
| **Pologne** | 1993 :01 | 24.011 | 203.258 | 2.622* | 16.428* | 1548.862* | 0.109** | 21.338** |
| **Turquie** | 1988 :01 | 21.340 | 226.663 | 0.764* | 1.493* | 36.632* | 0.092 | 13.441 |

*Moyen-orient et Afrique*

|  | Début | Moyenne | Ecart-Type | Skewness | Kurtosis | J.B. | 1ère auto. | Q(12) |
|---|---|---|---|---|---|---|---|---|
| **Afrique du Sud** | 1973 :01 | 9.214 | 101.134 | -0372* | 1.445* | 40.235* | 0.069 | 6.583 |
| **Egypte** | 1995 :01 | 3.466 | 100.263 | 1.188* | 2.790* | 56.564* | 0.225* | 16.411 |
| **Jordanie** | 1988 :01 | -1.713 | 53.755 | -0.139 | 1.428* | 16.335* | 0.141** | 10.947 |
| **Maroc** | 1995 :01 | 4.172 | 57.909 | 0.371 | 0.448 | 3.175 | 0.160** | 24.209** |
| **Tunisie** | 1998 :01 | 9.058 | 84.729 | 0.527** | 2.084* | 14.776* | -0.085 | 18.899 |

**Monde**

|  | Début | Moyenne | Ecart-Type | Skewness | Kurtosis | J.B. | 1ère auto. | Q(12) |
|---|---|---|---|---|---|---|---|---|
| **Monde** | 1973 :01 | 3.898 | 51.933 | -0.376* | 1.187* | 30.027* | 0.059 | 14.259 |

*\* significatif au seuil de 1%,  \*\* significatif au seuil de 5%, (1) centré sur 3, Q(12) test de Ljung-Box d'ordre 12 et J.B. test de normalité de Jarque-Bera.*

**Panel B- Indices des taux de change**

|  | Début | Moyenne | Ecart-Type | Skewness | Kurtosis | J.B. | 1ère auto. | Q(12) |
|---|---|---|---|---|---|---|---|---|
| **Pays développés** | 1973 :01 | -0.187 | 24.069 | 0.593* | 5.004* | 402.278* | 0.083 | 17.445 |
| **Pays émergents** | 1973 :01 | 0.531 | 14.462 | 1.025* | 4.452* | 364.453* | 0.184* | 46.340* |

*\* significatif au seuil de 1%,  \*\* significatif au seuil de 5%, (1) centré sur 3, Q(12) test de Ljung-Box d'ordre 12 et J.B. test de normalité de Jarque-Bera.*

**Panel C - Facteurs de l'intégration**

|  | Tous les pays | | Pays développés | | Pays émergents | |
|---|---|---|---|---|---|---|
| **Facteurs** | **Moyenne** | **Ecart-type** | **Moyenne** | **Ecart-type** | **Moyenne** | **Ecart-type** |
| **Degré d'ouverture commerciale (%)** | 63.21 | 67.04 | 65.35 | 68.53 | 61.08 | 58.87 |
| **Développement du marché boursier (%)** | 26.345 | 28.369 | 35.056 | 37.455 | 11.088 | 18.566 |
| **Actifs étrangers (%)** | 4.081 | 7.556 | 5.224 | 9.315 | 2.884 | 4.822 |
| **Actifs étrangers nets (%)** | 1.442 | 2.778 | 2.085 | 3.487 | 0.784 | 1.524 |
| **Volatilité des taux de change** | 0.272 | 0.273 | 0.366 | 0.278 | 0.186 | 0.241 |
| **Inflation (%)** | 8.475 | 14.030 | 5.657 | 5.139 | 11.363 | 18.844 |
| **Ecart des taux d'inflation (%)** | 2.849 | 14.033 | -0.198 | 5.293 | 6.019 | 18.789 |
| **Taux d'intérêt court (%)** | 10.120 | 12.031 | 7.601 | 4.436 | 12.918 | 15.685 |
| **Variation du taux d'intérêt court (%)** | -0.053 | 3.890 | -0.008 | 1.198 | -0.062 | 5.607 |
| **Prime de terme (%)** | 1.519 | 5.612 | 0.922 | 2.249 | 2.689 | 9.019 |
| **Ecart des taux d'intérêt courts (%)** | 4.216 | 11.447 | 1.100 | 3.664 | 8.186 | 17.119 |
| **Croissance de la production industrielle (%)** | 0.983 | 13.295 | 1.243 | 16.384 | 0.631 | 7.258 |
| **Ecart des taux de croissance (%)** | 0.096 | 1.584 | 0.120 | 1.956 | 0.060 | 0.864 |
| **Déficit courant (%)** | -2.138 | 8.501 | -3.178 | 6.539 | -1.100 | 9.981 |
| **Chômage (%)** | 7.544 | 3.392 | 7.487 | 2.579 | 7.729 | 5.235 |
| **Volatilité des exportations (en Log)** | 10.847 | 23.965 | 6.965 | 8.604 | 13.666 | 29.462 |
| **Taux d'intérêt mondial (%)** | 6.505 | 2.912 | - | - | - | - |
| **Variation du taux d'intérêt mondial (%)** | -0.011 | 0.452 | - | - | - | - |
| **Croissance de la production industrielle mondiale (%)** | 1.771 | 7.833 | - | - | - | - |
| **Risque pays** | 65.88 | 22.38 | 86.11 | 8.36 | 49.82 | 16.12 |

**Tableau 2 : Facteur de l'intégration internationale des marchés boursiers**

**Panel A- Modèle avec effets individuels**

| | Tous les pays | | | Pays développés | | | Pays émergents | | |
|---|---|---|---|---|---|---|---|---|---|
| **Facteurs** | $\kappa$ | Nbre obs. | Nbre pays | $\kappa$ | Nbre obs. | Nbre pays | $\kappa$ | Nbre obs. | Nbre pays |
| **Degré d'ouverture commerciale (%)** | 0.044*<br>*(0.016)* | 7482 | 30 | 0.039*<br>*(0.012)* | 3650 | 10 | 0.043*<br>*(0.017)* | 3832 | 20 |
| **Développement du marché boursier (%)** | 0.598<br>*(0.390)* | 7306 | 30 | 0.213<br>*(0.553)* | 3602 | 10 | 0.454***<br>*(0.246)* | 3704 | 20 |
| **Actifs étrangers (%)** | 2.504<br>*(2.781)* | 7026 | 30 | 2.317<br>*(5.114)* | 3433 | 10 | 2.874<br>*(2.205)* | 3593 | 20 |
| **Actifs étrangers nets (%)** | 0.026<br>*(0.098)* | 6257 | 28 | 15.135<br>*(47.144)* | 3094 | 10 | 0.066<br>*(0.118)* | 3163 | 18 |
| **Volatilité taux change** | -0.616***<br>*(0.363)* | 7088 | 29 | -0.486<br>*(0.736)* | 3276 | 9 | -1.571*<br>*(0.489)* | 3812 | 20 |
| **Inflation (%)** | 0.787<br>*(1.335)* | 6490 | 29 | -0.780<br>*(1.067)* | 3285 | 9 | 0.081<br>*(0.112)* | 3205 | 20 |
| **Ecart des taux d'inflation (%)** | 0.017<br>*(0.075)* | 6490 | 29 | 0.087<br>*(0.082)* | 3285 | 9 | 0.025<br>*(0.054)* | 3205 | 20 |
| **Taux d'intérêt court (%)** | 0.303<br>*(0.435)* | 6876 | 30 | -0.078<br>*(0.210)* | 3584 | 10 | 0.349<br>*(0.623)* | 3292 | 20 |
| **Variation du taux d'intérêt court (%)** | 3.409***<br>*(1.875)* | 6846 | 30 | 1.678<br>*(1.229)* | 3574 | 10 | 2.503<br>*(1.655)* | 3272 | 20 |
| **Prime de terme (%)** | 0.262<br>*(0.292)* | 5317 | 22 | 0.836<br>*(1.386)* | 3522 | 10 | 0.364<br>*(0.849)* | 1795 | 12 |
| **Ecart des taux d'intérêt courts (%)** | 0.525<br>*(0.714)* | 6856 | 30 | -0.283<br>*(0.748)* | 3584 | 10 | 0.048<br>*(0.089)* | 3272 | 20 |
| **Croissance de la production industrielle (%)** | 0.553<br>*(1.235)* | 5182 | 23 | -0.059<br>*(0.070)* | 2978 | 9 | 0.707<br>*(3.790)* | 2204 | 14 |
| **Ecart des taux de croissance (%)** | 48.254<br>*(65.337)* | 5182 | 23 | 5.870<br>*(7.158)* | 2978 | 9 | 17.65<br>*(64.790)* | 2204 | 14 |
| **Déficit courant (%)** | 0.064<br>*(0.126)* | 4667 | 21 | 0.058<br>*(0.506)* | 2333 | 8 | 0.071<br>*(0.380)* | 2334 | 13 |
| **Chômage (%)** | 0.140<br>*(0.327)* | 1700 | 13 | -1.074<br>*(0.739)* | 1302 | 8 | 0.163<br>*(0.439)* | 398 | 5 |
| **Volatilité des exportations (en Log)** | -0.785<br>*(1.005)* | 7469 | 30 | -0.254<br>*(0.558)* | 3650 | 10 | -0.025<br>*(0.115)* | 3819 | 20 |
| **Taux d'intérêt mondial (%)** | -0.100<br>*(0.074)* | 7482 | 30 | -0.126**<br>*(0.064)* | 3650 | 10 | -0.129<br>*(0.159)* | 3832 | 20 |
| **Variation du taux d'intérêt mondial (%)** | 59.805<br>*(64.276)* | 7482 | 30 | 2.005**<br>*(0.939)* | 3650 | 10 | 67.931<br>*(76.635)* | 3832 | 10 |
| **Croissance de la production industrielle mondiale (%)** | 0.433<br>*(0.664)* | 7482 | 30 | 0.405<br>*(0.312)* | 3650 | 10 | 0.604<br>*(0.883)* | 3832 | 20 |
| **Risque pays (log)** | 1.730***<br>*(1.012)* | 6442 | 30 | 1.526<br>*(2.413)* | 2850 | 10 | 1.964***<br>*(1.122)* | 3592 | 20 |

\* significatif au seuil de 1%, \*\* significatif au seuil de 5%, \*\*\* significatif au seuil de 10%, l'écart-type corrigé de l'autocorrélation-hétéroscédasticité (méthode de Newey et West (1987)) est reporté entre parenthèses.

**Panel B- Modèle sans effets individuels**

| | Tous les pays | | | Pays développés | | | Pays émergents | | |
|---|---|---|---|---|---|---|---|---|---|
| **Facteurs** | $\kappa$ | Nbre obs. | Nbre pays | $\kappa$ | Nbre obs. | Nbre pays | $\kappa$ | Nbre obs. | Nbre pays |
| **Degré d'ouverture commerciale (%)** | 0.046*<br>*(0.018)* | 7482 | 30 | 0.033*<br>*(0.014)* | 3650 | 10 | 0.053*<br>*(0.007)* | 3832 | 20 |
| **Développement du marché boursier (%)** | 0.621<br>*(0.387)* | 7306 | 30 | 0.388<br>*(0.733)* | 3602 | 10 | 0.356<br>*(0.235)* | 3704 | 20 |
| **Actifs étrangers (%)** | 2.547<br>*(1.863)* | 7026 | 30 | 1.903<br>*(4.358)* | 3433 | 10 | 3.453***<br>*(2.001)* | 3593 | 20 |
| **Actifs étrangers nets (%)** | 0.338<br>*(1.017)* | 6257 | 28 | 13.149<br>*(35.389)* | 3094 | 10 | 1.608<br>*(3.501)* | 3163 | 18 |
| **Volatilité taux change** | -0.524***<br>*(0.309)* | 7088 | 29 | -0.674<br>*(0.662)* | 3276 | 9 | -2.063*<br>*(0.371)* | 3812 | 20 |
| **Inflation (%)** | -0.021<br>*(0.157)* | 6490 | 29 | -0.730<br>*(1.675)* | 3285 | 9 | 1.679<br>*(1.580)* | 3205 | 20 |
| **Ecart des taux d'inflation (%)** | 0.156<br>*(0.207)* | 6490 | 29 | 0.118<br>*(0.604)* | 3285 | 9 | 0.078<br>*(0.093)* | 3205 | 20 |
| **Taux d'intérêt court (%)** | 0.073<br>*(0.060)* | 6876 | 30 | -0.068<br>*(0.177)* | 3584 | 10 | 0.061<br>*(0.332)* | 3292 | 20 |
| **Variation du taux d'intérêt court (%)** | 2.563<br>*(1.936)* | 6846 | 30 | 1.662<br>*(1.109)* | 3574 | 10 | 1.987<br>*(1.488)* | 3272 | 20 |
| **Prime de terme (%)** | 0.308<br>*(0.378)* | 5317 | 22 | 0.865<br>*(1.193)* | 3522 | 10 | 2.359<br>*(4.926)* | 1795 | 12 |
| **Ecart des taux d'intérêt courts (%)** | 0.609<br>*(1.411)* | 6856 | 30 | -0.236<br>*(0.708)* | 3584 | 10 | 0.029<br>*(0.109)* | 3272 | 20 |
| **Croissance de la production industrielle (%)** | 0.389<br>*(0.726)* | 5182 | 23 | 0.057<br>*(0.064)* | 2978 | 9 | 0.194***<br>*(0.114)* | 2204 | 14 |
| **Ecart des taux de croissance (%)** | 123.357<br>*(78.357)* | 5182 | 23 | 0.001<br>*(0.005)* | 2978 | 9 | 19.705<br>*(72.85)* | 2204 | 14 |
| **Déficit courant (%)** | 0.079<br>*(0.271)* | 4667 | 21 | 0.021<br>*(0.443)* | 2333 | 8 | 0.073<br>*(0.487)* | 2334 | 13 |
| **Chômage (%)** | -1.031***<br>*(0.590)* | 1700 | 13 | -0.869***<br>*(0.520)* | 1302 | 8 | 0.169<br>*(0.442)* | 398 | 20 |
| **Volatilité des exportations (en Log)** | -0.028<br>*(0.152)* | 7469 | 30 | -0.204<br>*(0.396)* | 3650 | 10 | 0.276<br>*(0.455)* | 3819 | 20 |
| **Taux d'intérêt mondial (%)** | -0.136<br>*(0.089)* | 7482 | 30 | -0.110**<br>*(0.054)* | 3650 | 10 | -0.203<br>*(0.188)* | 3832 | 20 |
| **Variation du taux d'intérêt mondial (%)** | 51.314<br>*(39.696)* | 7482 | 30 | 1.817***<br>*(1.021)* | 3650 | 10 | 81.051<br>*(83.902)* | 3832 | 10 |
| **Croissance de la production industrielle mondiale (%)** | 0.476<br>*(0.513)* | 7482 | 30 | 0.420***<br>*(0.248)* | 3650 | 10 | 0.772<br>*(1.008)* | 3832 | 12 |
| **Risque pays (log)** | 1.806<br>*(1.390)* | 6442 | 30 | 1.592<br>*(2.134)* | 2850 | 10 | 1.952<br>*(1.815)* | 3592 | 20 |

- significatif au seuil de 1%, ** significatif au seuil de 5%, *** significatif au seuil de 10%, l'écart-type corrigé de l'autocorrélation-hétéroscédasticité (méthode de Newey et West (1987)) est reporté entre parenthèses.